\documentclass[twocolumn,showpacs,prl]{revtex4}
\usepackage{amssymb}
\usepackage{graphicx}
\usepackage{dcolumn}
\usepackage{bm}
\usepackage{amsmath}

\begin{document}

\title{The Roton in a Bose-Einstein Condensate}
\author{J. Steinhauer$^1$}
\author{R. Ozeri$^2$}
\author{N. Katz$^2$}
\author{N. Davidson$^2$}
\affiliation{$^1$Department of Physics, Technion – Israel
Institute of Technology, Technion City, Haifa 32000, Israel}
\affiliation{$^2$Department of Physics of Complex Systems,
Weizmann Institute of Science, Rehovot 76100, Israel}

\begin{abstract}
The roton in a Bose-Einstein condensate is computed, both near and
far from a Feshbach resonance.  A low-density approximation is
made, allowing for an analytic result. A Monte Carlo calculation
shows that the roton is larger than predicted by the low-density
approximation, for the upper range of densities considered here.
The low-density approximation is applied to superfluid $^{4}$He,
roughly reproducing the results of previous Monte Carlo
calculations.
\end{abstract}

\maketitle

A roton, first predicted by Landau \cite{landau47} for superfluid
$^{4}$He, is an excitation in a Bose-Einstein condensed fluid,
characterized by a minimum in the excitation spectrum $\omega(k)$.
By Feynman's relation \cite{feynman54}, this minimum also
corresponds to a maximum in the static structure factor $S(k)$.
This peak in $S(k)$ occurs for excitations whose wavelength $2\pi
/k$ is equal to the characteristic wavelength of density
fluctuations in the ground-state wave function of the quantum
fluid.  The peak in $S(k)$ exceeds unity, the value for an
uncorrelated gas.

Feynman \cite{feynman54} described several possible microscopic
pictures of a roton in superfluid $^{4}$He.  He found that the
most likely description is that the roton is analogous to a single
atom moving through the condensate, with wave number $k$ close to
$2\pi/n^{-1/3}$, where $n^{-1/3}$ is the mean atomic spacing.

The roton in superfluid $^{4}$He was calculated by a Monte Carlo
technique \cite{mcmillan65,schiff67} using a many-body Jastrow
wave function \cite{jastrow55}, which is of the form
\begin{equation}\label{jastrow}
  \psi=\prod_{j>i=1}^{N}f(|\textbf{r}_{i}-\textbf{r}_{j}|),
\end{equation}
where the pair function $f(r)$ should be determined for the
quantum fluid under consideration.  The wave function
(\ref{jastrow}) deviates significantly from unity if any two atoms
become very close to one another.  For superfluid $^{4}$He, the
two-particle correlation function $g(r)$ for this wave function
has fluctuations with a preferential length scale of $n^{-1/3}$,
which results in a peak greater than unity in $S(k)$ near
$k\approx 2\pi /n^{-1/3}$, in agreement with Feynman's picture.

Both Feynman's result and the results of the Jastrow wave function
roughly agree with measurements \cite{henshaw60} of $S(k)$ and
$\omega(k)$.

The Jastrow wave function (\ref{jastrow}) has also been considered
for a low-density Bose-Einstein condensate (BEC)
\cite{mcmillan65}, and has been used to compute various properties
of a high-density BEC \cite{cowell02}.  We employ a Jastrow wave
function to compute the roton for a low-density BEC.  By an
analytic calculation, we find that the hard-sphere size $r_{o}$ of
the atoms determines the location of the roton, rather than
$n^{-1/3}$.  We see that Feynman's view that $n^{-1/3}$ is the
relevant length scale is the special case of high-density, for
which $r_{o} \approx n^{-1/3}$.

In a low-density BEC, Feynman's view of the roton as a single atom
moving through the condensate seems natural.  In general, the
excitation spectrum of a low-density BEC is of the Bogoliubov form
\cite{bogo47,steinhauer02}, which consists of phonons and
single-particle excitations.  We find that the roton occurs on the
single-particle part of the spectrum, at $k\approx 8/\pi a$, where
$a$ is the $s$-wave scattering length.

Although the general form of the wave function (\ref{jastrow}) can
be applied to both superfluid $^{4}$He and BEC, the wave function
is fundamentally different for these two quantum fluids.  For both
of these fluids, below some temperature $T_{s-wave}$, the thermal
de-Broglie wavelength $\lambda_{dB}$  becomes longer than the
characteristic length scale $R$ of the interparticle potential.
Below $T_{s-wave}$, all scattering processes except for $s$-wave
scattering become negligible.  Since a BEC is a dilute gas,
$n^{-1/3}$, which is typically about $1500\AA$, is much larger
than $R$.  Therefore, the critical temperature $T_{c}$ for quantum
degeneracy, at which $\lambda_{dB}$ becomes comparable to
$n^{-1/3}$, is much lower than $T_{s-wave}$. Therefore, only
$s$-wave scattering plays a role here for a BEC.  In contrast to
BEC, superfluid $^{4}$He is a relatively dense liquid, for which
$n^{-1/3}$ is comparable to $R$.  Therefore, the temperature
$T_{s-wave}$ also marks the transition $T_{c}$ to quantum
degeneracy. Even below $T_{c}$ for superfluid $^{4}$He, partial
waves other than the $s$-wave contribute to the wave function.

For many properties of a BEC, $a$ is sufficient to describe the
interparticle potential, and the form of the potential does not
play a role \cite{huang,dalfovo99}.  To quantitatively describe
the roton however, the $s$-wave scattering wave function must be
known to interparticle separations somewhat smaller than $R$,
where the details of the potential are relevant.

Rotons have been predicted in a BEC in the presence of optical
fields \cite{stamperkurn02,odell03}, as well as in a dipolar BEC
\cite{shlyapnikov03}. In this work, we find that a roton occurs in
an unperturbed BEC without any dipole interaction. We consider the
enhancement of the roton near a Feshbach resonance
\cite{feshbach,inouye}, but the physics is qualitatively the same
as in the unperturbed case.

When we refer to a roton, we are referring to a peak in $S(k)$,
rather than a minimum in $\omega(k)$.  By the Feynman relation,
the peak in $S(k)$ computed here for a BEC is not steep enough to
produce a minimum in $\omega(k)$, for the range of densities
considered here.

We use a low-density approximation \cite{jastrow55} to compute
$g(r)$ and the roton for a BEC with a positive scattering length.
The low-density approximation is compared to a Monte Carlo
calculation.

For a BEC of alkali atoms, the potential can be taken as the van
der Waals potential $-C_6/r^{6}$ for atomic separations $r$
greater than a few Angstroms \cite{flambaum93, leggett01}. For
this potential, $R$ is given by $(m C_6 / \hbar^{2})^{1/4}$, which
ranges from about $50$ to $100\AA$ \cite{leggett01,marinescu94}.
The wave function for $s$-wave scattering between two particles in
the limit of zero energy is given by $\psi_6(r)$, which is the
solution of the radial Schr\"{o}dinger equation with the van der
Walls potential. $\psi_6(r)$ can be written in terms of elliptic
integrals \cite{flambaum93}.  For $r>R$, this wave function is
very close in form to $1-a/r$ \cite{huang,comment1}. For the case
of $a\gg R$, $\psi_6(r)$ therefore obtains large negative values
for $r\ll a$ \cite{dalibard}.

We cannot use $\psi_6(r)$ as $f(r)$ in (\ref{jastrow}), because
even for $r\gg n^{-1/3}$, $\psi_6(r)$ is significantly less than
unity. This is non-physical, in the sense that no matter how large
the gas for fixed density, the value of the many-body wave
function (\ref{jastrow}) depends on the size of the gas.  To
account for many-body effects, we use the following pair function
which goes to unity for $r>n^{-1/3}$ \cite{leggett01,cowell02}.
\begin{equation}\label{fbec}
f(r)=
  \begin{cases}
    \psi_6(r) / \psi_6(n^{-1/3}) & (r\leq n^{-1/3}) \\
    1 & (r>n^{-1/3}).
  \end{cases}
\end{equation}
To compute (\ref{fbec}), $R$ is needed.  Throughout this work, we
use $R=0.05$ $n^{-1/3}$. This is a typical experimental value, and
the results here are rather insensitive to $R$.

Equation (\ref{jastrow}) with (\ref{fbec}) is shown schematically
in Fig. 1. To aid in visualization, Fig. 1 shows the wave function
squared in one dimension, as a function of the position $x_{1}$ of
atom number $1$. The positions of all of the other atoms, such as
atoms $b$ through $e$, are fixed.  As long as atom number $1$ is
far from the other atoms, the wave function has the constant value
$\psi_{o}$. This value is determined by the positions of the atoms
other than $x_{1}$.  The rapid oscillations in $\psi_6(r)$ for $r
< R$ appear in Fig. 1 as dark vertical bands, when atom number 1
is very close to another atom.

The correlation function $g(r)$ gives the unconditional
probability of two atoms being at a distance $r$.  $g(r)$ is
related to the pair function $f(r)$ by \cite{jastrow55}
\begin{equation}\label{gofr}
  g(|\textbf{r}_{1}-\textbf{r}_{2}|)=V^2\frac{\int d\textbf{r}_{3}...d\textbf{r}_{N}\prod\limits_{j>i=1}^{N}f^{2}(|\textbf{r}_{i}-\textbf{r}_{j}|)}{\int
  d\textbf{r}_{1}...d\textbf{r}_{N}\prod\limits_{j>i=1}^{N}f^{2}(|\textbf{r}_{i}-\textbf{r}_{j}|)},
\end{equation}
where the integrals are over the volume $V$.  (\ref{gofr}) can be
evaluated starting with the first integral in the denominator,
$\int
d\textbf{r}_{1}\prod\limits_{j>1}^{N}f^{2}(|\textbf{r}_{1}-\textbf{r}_{j}|)$.
This can be visualized as the integral over the function shown in
Fig. 1 for one dimension.  Neglecting three-body interactions
greatly simplifies this integral.  Three-body interactions are
rare for small values of the gas parameter $na^{3}$, assuming that
the range of three-body interactions is of the same order of
magnitude as the range of two-body interactions. A three-body
interaction is represented in Fig. 1 by the points $x_{c}$ and
$x_{d}$, where atoms $1$, $c$, and $d$ interact. Neglecting such
interactions, the integral is a function of the volume $v$
indicated by the shaded region in Fig. 1. The integral is then
given by $V-(N-1)v$, where $v=V-\int d\textbf{r}f^{2}(r)$. This
result is independent of the positions of the $\textbf{r}_{j}$.
\begin{figure}[h]
\begin{center}
\includegraphics[width=3.3in]{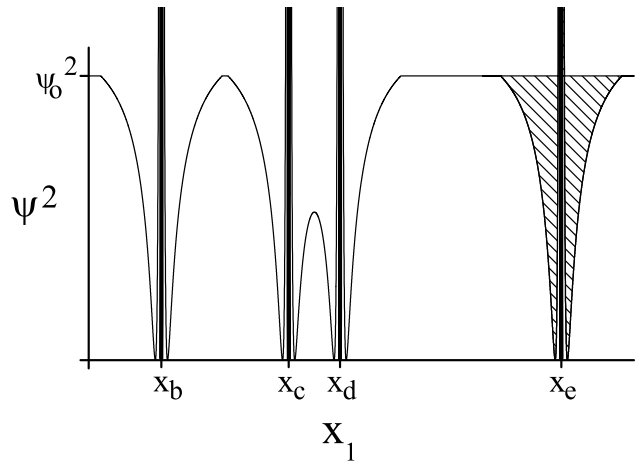}
\end{center}
\caption{Schematic one-dimensional representation of the Jastrow
wave function.  The dependence on the single dimension $x_{1}$ is
shown. The overall scaling of the curve is given by the factors in
the wave function not involving $x_{1}$.  The labeled values of
$x_{1}$ correspond to the positions of atoms other than atom $1$.
$x_{1}$ varies over the volume $V$.}
\end{figure}

Evaluating all of the integrals in (\ref{gofr}) similarly to the
first yields $g(|\textbf{r}_{1}-\textbf{r}_{2}|)\approx
V^2f^{2}(|\textbf{r}_{1}-\textbf{r}_{2}|)[(V-v)V]^{-1}$. Since
$V\gg v$, we obtain the result of the low-density approximation
\cite{jastrow55}
\begin{equation}\label{gapprox}
  g(r)\approx f^{2}(r).
\end{equation}
The result (\ref{gapprox}) with (\ref{fbec}) is indicated by the
solid curve in Fig. 2 for $na^{3}=2.2\times 10^{-4}$.  This value
of $na^3$ is an order of magnitude greater than typical
experimental values without a Feshbach resonance. The result for
$na^{3}=0.011$ is also shown in the figure.

While (\ref{gapprox}) is useful for obtaining analytic results, a
more accurate computation can be made by the Monte Carlo technique
described in Ref. \cite{mcmillan65}.  This technique effectively
evaluates (\ref{gofr}) by using a Metropolis algorithm to randomly
choose configurations of the $N$ atoms ($N=100$ here), according
to the probability distribution given by (\ref{jastrow}) with
(\ref{fbec}), and computing the distribution of distances between
the atoms, with periodic boundary conditions. This distribution,
averaged over many likely configurations, is proportional to
$r^{2}g(r)$.  The result of the Monte Carlo computation is shown
in Fig. 2 for $na^3=2.2\times 10^{-4}$ and 0.011.  These results
were obtained with $9\times 10^8$ and $2\times 10^7$ iterations,
respectively.
\begin{figure}[h]
\begin{center}
\includegraphics[width=3.3in]{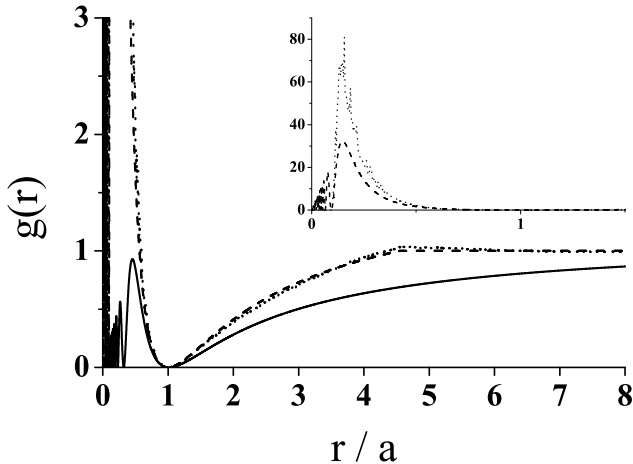}
\end{center}
\caption{The two-particle correlation function for a BEC. The
solid and dashed curves are the low-density approximation
(\ref{gapprox}) with (\ref{fbec}), with $na^3=2.2\times 10^{-4}$
and 0.011, respectively. The dotted curve is the Monte Carlo
result for $na^3=0.011$.  For $na^3=2.2\times 10^{-4}$, the Monte
Carlo result is indistinguishable from the solid curve.  The inset
shows the curves for $na^3=0.011$ only.}
\end{figure}

The small oscillations for small $r$ shown in Fig. 2 are
negligible in the computation of the static structure factor.
These are the oscillations for r values less than that of the
first large peak below $r/a=1$, in the solid and dashed curves.
To save computing time these oscillations are not included in
$f^2(r)$ in the Monte Carlo computation.

For $a\ge 1.3 R$ (corresponding to $na^3>2.7\times 10^{-4}$ in
this work), The peak in $f^2(r)$ located at $r<R$ is greater than
unity. The dashed line of the inset of Fig. 2 shows an example of
such a peak, for $a = 4.4R$. Because this peak is greater than
unity, a cluster of atoms is the most likely configuration. Since
we are interested in the un-clustered phase, we do not let the
Monte Carlo computation proceed long enough for the clear
transition to clusters to occur.

The static structure factor $S(k)$ is given by $1+n\int
[g(r)-1]e^{i\textbf{k}\cdot \textbf{r}}d\textbf{r}$, which in
general can be written
\begin{equation}\label{sofk}
  S(k)=1+4\pi n
  \int\limits_{0}^{\infty}drr^{2}[g(r)-1]\frac{sin(kr)}{kr}.
\end{equation}
We use (\ref{sofk}) to compute $S(k)$ for the low-density
approximation (\ref{gapprox}) with (\ref{fbec}), as indicated in
Fig. 3. The height $S(k_{r})$ and location $k_{r}$ of the roton
are indicated by the solid curves of Fig. 4. $S(k)$ in Fig. 3 has
several maxima, the tallest of which is taken as the roton. As
$na^3$ is varied, the peak which is taken as the roton varies,
resulting in the jagged appearance of the solid curves of Fig. 4.

For the Monte Carlo computation, $S(k)$ is found by inserting
$g(r)$ such as is shown in Fig. 2 into (\ref{sofk}). The results
are indicated in Fig. 3 by the dash dotted and solid curves.  For
small $na^3$, the height and location of the roton are seen in
Fig. 4 to be the same for the low-density approximation and the
Monte Carlo result.  The low-density approximation is therefore
valid for small $na^3$. For the larger values of $na^3$, the
height of the roton in the Monte Carlo calculation is larger than
that of the low-density approximation, as seen in Fig. 4a.

By applying the low-density approximation to the limit of $a\gg
R$, we can obtain an analytic expression for $S(k)$ for $R\ll a\ll
n^{-1/3}$.  In this range, $f(r)$ can be taken as $1-a/r$. By
(\ref{gapprox}) and (\ref{sofk}),
\begin{equation}\label{sofkbec}
S(k)=1+4\pi na^3[\pi(ka)^{-1}/2-2(ka)^{-2}].
\end{equation}
This small-R limit is indicated by the dotted curve in Fig. 3. The
height of the roton in this limit is $S(k_{r})=1+\pi^3na^3/8$, and
the location is given by $k_{r}=8/\pi a$.  These values are
indicated by the dashed curves of Fig. 4, which as expected, agree
with the low-density approximation (solid curves) for $a\gg R$,
where $nR^3$ is indicated by the dotted lines.  As seen in the
above expression for $k_r$, the location of the roton is
determined by $a$, rather than by $n^{-1/3}$.

\begin{figure}[h]
\begin{center}
\includegraphics[width=3.3in]{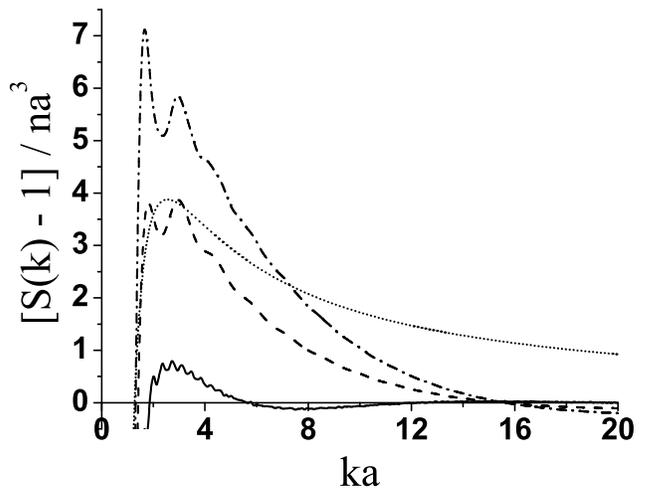}
\end{center}
\caption{The roton in a BEC, computed by (\ref{sofk}).  The dash
dotted and solid curves are the Monte Carlo results for
$na^3=0.011$ and $2.2\times 10^{-4}$ respectively.  The result of
the low-density approximation (\ref{gapprox}) with (\ref{fbec})
for $na^3=0.011$ is indicated by the dashed curve.  The result of
the low-density approximation for $na^3=2.2\times 10^{-4}$ is
indistinguishable from the solid curve. The dotted curve is the
small-R limit, given by (\ref{sofkbec}).}
\end{figure}

For $na^3=0.011$, the height of the roton given by the Monte Carlo
calculation is $S(k_r)=1.08$, as shown in Figs. 3 and 4a.
Measuring this $8\%$ effect could be experimentally feasible.
$na^3=0.011$ could be attained by a Feshbach resonance
\cite{comment1}.  It should be noted though, that for this
relatively large value of $na^3$, the form of (\ref{fbec}) is only
approximate, so the expressions for the height and location of the
roton should be considered as estimates only.

Due to phonons, the true $S(k)$ is proportional to $k$ for
$k\lesssim \xi^{-1}$, where $\xi^{-1}=a^{-1}\sqrt{8\pi na^3}$ is
the inverse healing length. The curves shown in Fig. 3 do not show
this linear behavior for small $k$ because the wave function
(\ref{jastrow}) with (\ref{fbec}) does not have the long-range
correlations of a phonon \cite{mcmillan65,schiff67}.

For superfluid $^{4}$He the three inverse length scales
$2\pi/n^{-1/3}$, $k_{r}$, and $\xi^{-1}$, are roughly equal. This
is also true for a strong roton in a BEC.  As $na^3$ increases,
both $k_{r}$ and $\xi^{-1}$ approach $2\pi/n^{-1/3}$. More
precisely, the ratios of $\xi^{-1}$ and $k_{r}$ to $2\pi/n^{-1/3}$
are $\sqrt{2/\pi}(na^3)^{1/6}$ and approximately
$4/\pi^2(na^3)^{-1/3}$, respectively.  The latter ratio implies
that an appropriate measurement system for measuring a roton
should be able to probe wavelengths somewhat shorter than
$n^{-1/3}$.
\begin{figure}[h]
\begin{center}
\includegraphics[width=3.3in]{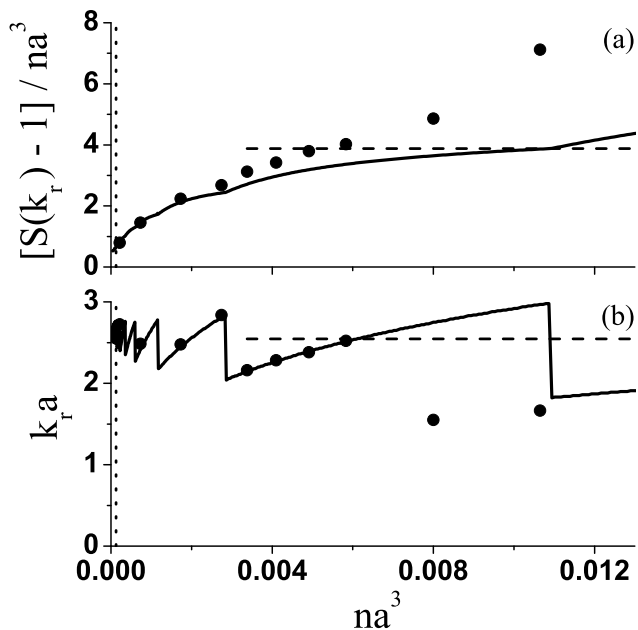}
\end{center}
\caption{The height (a) and location (b) of the roton in a BEC, as
a function of $na^3$.  The circles are the Monte Carlo result. The
solid curves are the low-density approximation.  The dashed curves
are the analytic small-R limit.  The dotted lines indicate
$nR^3$.}
\end{figure}

As a demonstration of the validity of the low-density
approximation (\ref{gapprox}), we use (\ref{gapprox}) to compute
various properties of superfluid $^{4}$He found in
\cite{mcmillan65} and \cite{schiff67} by means of Monte Carlo
calculations.  For superfluid $^{4}$He, $nb^3\approx 0.4$ where
$b$ is the hard-sphere size. Therefore, this is a stringent test
of the low-density approximation.  We find that the location of
the roton for superfluid $^{4}$He is given by $k_{r}=5.3/b$, which
differs from the Monte Carlo result by 6\%
\cite{mcmillan65,schiff67}.  We find that the height of the roton
for superfluid $^{4}$He is $S(k_{r})\approx 1+0.3nb^{3}$, which
gives a roton of $S(k_{r})=1.1$, compared to $S(k_{r})=1.2$ from
the Monte Carlo calculation \cite{mcmillan65,schiff67}, and
$S(k_{r})=1.5$ from experiment \cite{henshaw60}.  The
order-of-magnitude agreement between the low-density approximation
and the Monte Carlo result for superfluid $^{4}$He suggests that
the low-density approximation preserves the essence of
(\ref{gofr}).

In conclusion, we find the height $S(k_{r})$ and location $k_{r}$
of a roton in a BEC, for a range of densities.  A low-density
approximation is compared to a Monte Carlo calculation. The values
of $S(k_{r})$ and $k_{r}$ given by the two methods agree for the
lowest densities.  For higher densities, the Monte Carlo
calculation predicts an enhancement in $S(k_{r})$ of almost a
factor of 2 over the low-density approximation.

In contrast to the Monte Carlo calculation for superfluid
$^{4}$He, the small-R limit gives explicit expressions for the
height and location of the roton.

We thank Servaas Kokkelmans, Ady Stern, Yoseph Imry, Daniel
Kandel, Johnny Vogels, and Ananth Chikkatur for helpful
discussions. This work was supported by the Israel Science
Foundation.  J. S. is a Landau Fellow, supported by the Taub and
Shalom Foundations.

\end{document}